\documentclass[manuscript]{aastex}

\begin{document}

\title{ASCA Observation of the quiescent X-ray counterpart to SGR1627-41}

\author{K. Hurley\altaffilmark{1}, T. Strohmayer\altaffilmark{2},
P. Li\altaffilmark{1}, C. Kouveliotou\altaffilmark{3}, P. Woods\altaffilmark{4},
J. van Paradijs\altaffilmark{5}, T. Murakami\altaffilmark{6},
D. Hartmann\altaffilmark{7}, I. Smith\altaffilmark{8}, M. Ando\altaffilmark{6},
A. Yoshida\altaffilmark{9}, and M. Sugizaki\altaffilmark{10}}

\email{khurley@sunspot.ssl.berkeley.edu}
\altaffiltext{1}{University of California, Berkeley, Space Sciences Laboratory,
Berkeley, CA 94720-7450}
\altaffiltext{2}{NASA Goddard Space Flight Center, Greenbelt, MD 20771}
\altaffiltext{3}{Universities Space Research Association at NASA Marshall Space Flight Center, 
ES-84, Huntsville AL 35812}
\altaffiltext{4}{NASA/Marshall Space Flight Center, ES-84, Huntsville, AL 35812;
University of Alabama in Huntsville, AL}
\altaffiltext{5}{University of Alabama in Huntsville, AL 35899; Astronomical Institute `Anton Pannekoek', University of Amsterdam, The Netherlands}
\altaffiltext{6}{ISAS, 3-1-1 Yoshinodai, Sagamihara, Kanagawa, Japan 229}
\altaffiltext{7}{Clemson University, Department of Physics and Astronomy, Clemson SC, 
29634-0978}
\altaffiltext{8}{Department of Space Physics and Astronomy, Rice University, MS-108, 6100 South
Main, Houston, TX 77005-1892}
\altaffiltext{9}{Institute of Chemical and Physical Research (RIKEN), 2-1 Hirosawa, Wako,
Saitama 351-0198 Japan}
\altaffiltext{10}{Space Utilization Research Program, Tsukuba Space Center, National
Space Development Agency of Japan, 2-1-1 Sengen, Tsukuba, Ibaraki, 305-8505 Japan}

\begin{abstract}

We present a 2 - 10 keV ASCA observation of the field around the soft gamma
repeater SGR1627-41.  A quiescent X-ray source was detected in this observation 
whose position was consistent both with that of a recently discovered
\it BeppoSAX \rm X-ray source and with
the Interplanetary Network localization for this SGR.  
In 2 - 10 keV X-rays, the 
spectrum of the X-ray source may be fit equally well by a power law, blackbody, or bremsstrahlung function, with unabsorbed
 flux  $\approx \rm5 x 10^{-12} erg \ cm^{-2}\ s^{-1}$.  
We do not confirm a continuation of a fading trend in the flux, 
and we find no evidence for periodicity, both noted in the earlier 
\it BeppoSAX \rm observations.

\end{abstract}


\section{Introduction}

SGR1627-41, the fourth Soft Gamma Repeater, was discovered in a series of observations
between 1998 June and 1998 July
by the Burst and Transient Source Experiment (BATSE) aboard the \it Compton Gamma-Ray Observatory
\rm (Woods et al. 1999),
the Gamma-Ray Burst Experiment aboard \it Ulysses \rm (Hurley et al. 1999), the KONUS experiment aboard
the \it Wind \rm spacecraft (Mazets et al. 1999), and the All Sky Monitor on the 
\it Rossi X-Ray Timing Explorer \rm (Smith et al. 1999).  The Interplanetary Network (IPN) error
box for SGR1627-41 passes through the Galactic supernova remnant (SNR) G337.0-0.1 (Hurley et al.
1999).  Two \it BeppoSAX \rm observations of this region on 1998 August 7
and 1998 September 16 revealed an X-ray source, most
likely a neutron star, whose
position was consistent with that of both the IPN localization and the SNR; the source
displayed a possible periodicity of 6.41 s (chance probability $\rm 6
\times 10^{-3}$, based on a limited number of trials: Woods et al. 1999).

Based on BATSE observations of the SGR in outburst, Woods et al. (1999) estimated
that the neutron star magnetic field strength was $\rm \gtrsim 5 \times 10^{14} Gauss$.
These properties would make the SGR counterpart a magnetar, an object in which magnetic
energy dominates all other sources of energy, including rotation (Thompson and Duncan,
1995, 1996).  Evidence that other
SGRs are also magnetars has been presented (Kouveliotou et al. 1998, 1999; but see also
Marsden, Rothschild \& Lingenfelter 1999).  In those cases, decisive evidence for the
magnetic field strength came from the spindown rates, and their interpretation as
dipole radiation.  In an attempt to confirm the periodicity of SGR1627-41 and
measure its spindown rate, we observed the source with the \it Advanced Satellite
for Cosmology and Astrophysics \rm (ASCA).

\section{ASCA Observations}

The ASCA observation took place between 1999 February 26 and 1999 February 28.  
The nominal pointing direction was $\rm \alpha(2000)=16^{\rm h} 36^{\rm m} 14^{\rm s},
\delta(2000)=-47^{\rm o} 32 \arcmin 31 \arcsec$, and the approximate exposures were 
72.7 ks for the SIS and 78.4 ks for the GIS.  We used the standard screening criteria for such parameters as Earth elevation angle, South Atlantic
Anomaly, and cutoff rigidity to extract photons, as explained in the ASCA Data Reduction Guide, Version 2\footnote{http://heasarc.gsfc.nasa.gov/docs/asca/abc/abc.html}.
No bursts from the source
were observed by \it Ulysses \rm, BATSE, or ASCA during the observation 
(the last burst from SGR1627-41 was observed in 1998 August).
Using the Ximage source detection tool, a quiescent source was detected at $\rm \alpha(2000)= 16^{\rm h} 35^{\rm m} 46.41^{\rm s},
\delta(2000)=-47^{\rm o} 35 \arcmin 13.1 \arcsec$ with a 3 $\sigma$ error radius 55 $\arcsec$,
consistent with the
1 $\arcmin$ radius error circle of the \it BeppoSAX \rm source (figure 1).  Approximately 3800 net counts were detected, versus $\approx$ 2850 in the 
first \it BeppoSAX \rm observation.  Two other sources were detected in this observation 
(one is visible in figure 1),
but neither had a position consistent with either the IPN annulus or G337.0-0.1.
Assuming that the ASCA, \it BeppoSAX \rm, and SGR sources are the same object, the most likely
position of the SGR is around the intersection of the IPN annulus with this new error circle,
at  $\rm \alpha(2000)= 16^{\rm h} 35^{\rm m} 52^{\rm s},
\delta(2000)=-47^{\rm o} 35 \arcmin 14 \arcsec$.

The region used for spectral analysis consisted of a 105 $\arcsec$
radius circle centered at the source position; background was taken from the
same observation, using a
similar circle at a region where no source was present, as determined
by Ximage.  
Spectral fitting to the GIS2 and GIS3 data was done using XSPEC and three trial functions: blackbody, thermal
bremsstrahlung, and a power law, all with absorption.  These results are reported in
table 1, along with the earlier \it BeppoSAX \rm results.    There is no clear preference for any of these models, but we adopt the power law fit for further discussion.  

To search for periodicity, barycentric light curves were constructed with 0.125 s binning from
the sum of the GIS2 and GIS3 data, by extracting $\approx$ 1 - 10 keV counts from a 4 $\arcmin$ radius circular region around the
source, and an FFT was performed (figure 2).  The most prominent peak in the power spectrum was at 0.10821 Hz (significance 0.12).  The 90\% confidence upper limit to the power of any signal in the spectrum with period between 0.01 and 1 Hz is
$\approx$ 3\% (rms).  Woods et al. (1999) found a 6.413183 s period in the first of their two
\it BeppoSAX \rm observations, with an rms pulse fraction of 10\% $\pm$ 2.6\%.  The 
upper limit to the signal power at this period from the folded ASCA light curve is
1.8\% (rms).   This limit would be appropriate only if the spindown rate were zero; 
if the quiescent counterpart to SGR1627-41 is characterized by a rate $\rm \approx
10^{-10} s s^{-1}$, as is the case for SGR1900+14, the period could have changed
by as much as 0.0015 s between the \it BeppoSAX \rm and ASCA observations, and the 3\%
upper limit would be the appropriate one.  We can also state with varying degrees of
confidence that no significant periodicities exist between 0.001 and 0.01 Hz, although
in this range the period search is dominated by windowing effects from the data gaps
and non-uniform sampling.
 
\section{Discussion and Conclusion}

The earlier \it BeppoSAX \rm observations of the quiescent counterpart of SGR1627-41
indicated a fading trend, significant at the $\rm \approx 5.9 \sigma$ level in
the raw data, over the $\approx$ 5 week period between the two
pointings (Woods et al. 1999, and table 1).  The unabsorbed 2-10 keV flux found
in the ASCA observation reported here is consistent with that found in the
second \it BeppoSAX \rm observation, and therefore indicates that this trend 
did not continue.  We have checked this conclusion in two ways.
First, we performed
a joint fit to the \it BeppoSAX \rm and ASCA observations, and found that they could
be described well by a single power law with no change in normalization.   Second,
we calculated the unabsorbed 2-10 keV source flux fixing the best fit power law
index and $\rm N_H$ to those found in the \it BeppoSAX \rm observations, and confirmed that it agreed
with the \it BeppoSAX \rm flux.

Relatively little is known about the mechanisms for variability in the
quiescent soft X-ray counterparts to SGRs.  Since variability in the 
quiescent emission of SGR1900+14 has definitely been observed (Kouveliotou et al. 1999,
Murakami et al. 1999),
it seems plausible that the quiescent steady emission from SGR1627-41, varying 
at an earlier time, could have ceased to vary by the time of the observations reported
here.  It could also be argued that the periodic quiescent emission originated
on a cooling hot spot on the neutron star surface, and
that it became undetectable by the time of the ASCA observations.
In any case, however, compelling evidence that SGR1627-41 is a magnetar must await an 
unambiguous detection of the periodicity and a measurement of the spindown rate.   
Short \it Chandra \rm observations could resolve this and also determine the precise
source position.

\acknowledgments
KH and PL are grateful to NASA for support under the ASCA AO-7 
Guest Investigator Program.

\newpage
\figcaption{Two IPN annuli from Hurley et al. (1999), superimposed on the 843 MHz radio contours of
G337.0-0.1 from Whiteoak and Green (1996).  The \it BeppoSAX \rm error circle from Woods et al. (1999)
and the ASCA error circle from the present observation are indicated.  A second
ASCA source is visible in the upper right hand corner.\label{fig1}
}

\figcaption{Power spectrum of the 2 - 10 keV soft X-ray source associated with
SGR1627-41.    \label{fig2}
}

\newpage

\begin{deluxetable}{cccc}
\rotate
\tablecaption{\it Spectral fits to the soft X-ray counterpart of SGR1627-41}
\tablehead{
\colhead{}& \colhead{Power law}  & \colhead{Blackbody} & \colhead{Bremsstrahlung}
}
\startdata
\cutinhead{ASCA}
Index              & -2.2 $\pm$ 0.3 &                 &                 \\
kT, keV            &                  & 1.3 $\pm$ 0.1 & 6.5 $\pm$ 2 \\
Reduced $\chi^2$, 89 d.o.f.   & 0.90             & 0.91            & 0.86            \\
$\rm Column \, density, N_H,$ \\
$\rm 10^{22} cm^{-2}$
                   & $\rm 6.9 \pm 1 $
                                      & $\rm 3.2 \pm 0.7 $
                                                        & $\rm 5.9 \pm 0.8$ \\
Unabsorbed 2-10 keV flux, \\ 
$\rm erg \ cm^{-2} \ s^{-1}$ 
                   & $\rm (5.1 \pm 0.4) \times 10^{-12}$
                                      & $\rm (3.5 \pm 0.3) \times 10^{-12}$ 
                                                        & $\rm (4.6 \pm 0.4) \times 10^{-12}$ \\
\cutinhead{\it BeppoSAX\tablenotemark{1} \rm}

Index              & -2.5 $\pm$ 0.2   &                 &                \\
$\rm Column \, density, N_H, 10^{22} cm^{-2}$
                   & $\rm 7.7 \pm 0.8 $ &               &                \\
2-10 keV flux, 
$\rm erg \ cm^{-2} \ s^{-1}$, \\
1998 August 7 
                   & $\rm (6.7 \pm 0.3) \times 10^{-12}$ &         &     \\
2-10 keV flux,
$\rm erg \ cm^{-2} \ s^{-1}$, \\
1998 September 16 
                   & $\rm (5.2 \pm 0.4) \times 10^{-12}$ &         &      \\

\enddata
\tablenotetext{1}{Woods et al. 1999}
\end{deluxetable}

\end{document}